\documentclass[11pt,a4paper]{article}
\usepackage{jheppub} 
\usepackage{amsmath,amssymb,amsfonts}
\usepackage{physics}
\usepackage{xcolor}
\usepackage{cleveref}
\usepackage{hyperref}
\usepackage{tikz}
\usetikzlibrary{decorations.markings, arrows.meta, positioning}
\usepackage{orcidlink}

\newcommand{\Gn}{ {G_{N}}}

\title{Nonperturbative double copy:\\
 worldline instantons, color thermality, and backreaction}

\author[a,b]{John Joseph M. Carrasco\orcidlink{0000-0002-4499-8488},}
\author[a,b]{Yaxi Chen\orcidlink{0009-0003-4625-994X},}
\author[a]{Nicolas H. Pavao\orcidlink{0000-0002-9817-8266},}
\author[a]{Aslan Seifi\orcidlink{0009-0007-2322-5842}}

\affiliation[a]{The Amplitudes and Insights Group, Department of Physics \& Astronomy, Northwestern University, Evanston, Illinois 60208, USA}
\affiliation[b]{Center for Interdisciplinary Exploration and Research in Astrophysics (CIERA), Northwestern University, 1800 Sherman Ave, Evanston, IL 60201, USA}

\abstract{
We present a first-principles, non-perturbative worldline instanton analysis of vacuum decay in the non-abelian Yang-Mills root of a Schwarzschild background. We recover the gauge theory color-thermal spectrum as a topological winding mode. The double copy maps vacuum response from the gauge theory directly to the gravity theory.  Furthermore, the decay exponent acquires a universal quadratic correction from color charge conservation, showing that the double copy correctly captures the non-linear backreaction as required for unitarity.  
}

\begin{document}

\maketitle
\flushbottom
\newpage

\section{Introduction}
\label{sec:intro}

The double copy has emerged as a framework that structurally unifies theories by identifying shared quantum field-theoretic building blocks, organized by their little-group representations. This allows us to express gravitational interactions in terms of gauge theory interactions order by order in the coupling.   Originally discovered in the context of perturbative scattering amplitudes~\cite{Bern:2008qj, Bern:2010ue}, building on the venerable open-closed string factorization relations of Kawai, Lewellen and Tye~\cite{Kawai:1985xq} as well as their field theory limits~\cite{Berends:1988zp,Bern:1998sv}, the double copy replaces the color factors of Yang-Mills theory with kinematic numerators to generate amplitudes in quantum gravity.  More broadly, this suggests that Yang-Mills emerges as the interplay between two dual algebras --- one governing its gauge group, and one governing its kinematics.  Gravity allows us to experience the interplay of the kinematic algebra with itself.

Subsequent developments have shown that this structure extends beyond perturbation theory. Classical Double Copy~\cite{Monteiro:2014cda, Luna:2015paa, White:2024pve} establishes that certain classes of exact solutions to Einstein's equations, such as the Schwarzschild black hole, possess a specific gauge-theoretic root: a single-copy field configuration whose kinematic data generates the full spacetime curvature upon double copy. 

Recently, this correspondence has been applied to the quantum phenomenon of particle production. In a remarkably sharp application of modern amplitudes methods, Hawking's  radiation spectrum of a Schwarzschild black hole~\cite{Hawking:1975vcx} was rederived~\cite{Aoude:2024sve} by considering the eikonal scattering of probe particles in a Vaidya background, resummed via Lippmann-Schwinger.  The dual gauge theory calculation has demonstrated that the radiation spectrum from an SU($N_c$) root-Hawking (root-Vaidya) background (the Yang-Mills single copy of a collapsing shell) exhibits its own  thermal character~\cite{Carrasco:2025bgu}.  This is in striking contrast to complementary analyses in the abelian limit,
where no event by event thermality is found: the eikonal scattering treatment of
Ref.~\cite{Aoude:2025jvt} and the effective-action analysis of Ref.~\cite{Ilderton:2025aql}.

In the SU($N_c$) case  the differential spectrum is thermal, not in energy, but in the distribution of radiated color charge eigenvalue $\lambda$.  
 This establishes color thermality as an on-shell, real-time, intrinsic property of the S-matrix of the root theory.  
A fundamental question then remains: what is the non-perturbative semiclassical origin of this thermality? 
Does it arise from a genuine Euclidean saddle in the gauge theory, directly dual to the geometric origin of Hawking radiation at the horizon?

In this paper, we answer this question by analyzing the vacuum stability of the Yang-Mills root using the worldline instanton formalism~\cite{Affleck:1981bma,Dunne:2005sx}.  
This method provides a direct spacetime description of vacuum decay, identifying the semiclassical saddle points that mediate pair production.  
Equivalence to the real-time Bogoliubov description of Schwinger production was established in a series of works~\cite{Kim:2000un,Dunne:2006st,Dunne:2010zz}.
We demonstrate that the thermal spectrum found in scattering amplitude is enforced by a precise Euclidean saddle: a topological winding of the worldline around the color source.  
The resulting vacuum decay rate is governed by a universal color temperature
\begin{equation}
T_c \propto (gQ)^{-1},
\end{equation}
determined solely by the large background charge $Q$.

Throughout this work, we treat the Yang-Mills root as a macroscopic, coherent color source with a fixed orientation in color space.
This should be understood as a controlled semiclassical sector rather than a microscopic description of the source.
Our analysis applies in the regime of large charge $Q\gg 1$, where the background field outside the source is well-approximated 
by a classical configuration and vacuum decay is exponentially suppressed. In this ``cold'' limit, emission events are rare and 
separated by long intervals during which the source evolves adiabatically.  The observables we compute are dictated  only by the global analytic structure of the long-range field and therefore  insensitive to any short-range dynamics near (or within) the background source.

We note that related ideas connecting macroscopic, high-occupancy gauge-field
configurations to black-hole-like behavior in gravity have been explored from a
different complementary perspective, notably in the correspondence between color glass
condensates and graviton condensates~\cite{Dvali:2021ooc}.
There, the emphasis is on entropy bounds, unitarization, and the collective
microstate structure of saturated systems --- exploring classicality from the perspective of large occupation number.   In contrast, our analysis is agnostic as to the microphysical realization of the background gauge  field, instead focusing on universal features of vacuum response. 

The efficacy of our analysis relies on the unique structure of the Kerr-Schild background.  It is well established that the Kerr-Schild metric represents a kinematic regime where the infinite tower of graviton self-interactions collapses into a single linear propagator structure~\cite{Duff:1973zz}.  This algebraic simplification has a precise dual in the gauge theory: the abelianization of the Yang-Mills root.

The same fixed color alignment that linearizes the field equations also trivializes the path-ordering of the Wilson loop.  This allows us to reduce the generally intractable non-abelian worldline path integral to a solvable mechanics problem.  Crucially, however, the result retains the imprint of the non-abelian algebra: while the path integral factorizes, the resulting effective temperature and backreaction are distinct from the abelian theory.

A key result of this work is the clarification of the relationship between Schwinger pair production and Hawking radiation.  It is straightforward to see that  Schwinger provides a gauge-theoretic analog to Hawking radiation via the Unruh effect~\cite{Brout:1995rd}.  We demonstrate where this analogy breaks down in the abelian limit.  In Quantum Electrodynamics (QED), the effective temperature depends inversely on the probe mass ($T \sim 1/m$), violating the universality required by the Equivalence Principle ($T \sim 1/M$).

We show in \Cref{sec:nonabelian} that the non-abelian root cures this defect.  In the Yang-Mills instanton, the color charge eigenvalue $\lambda$ determines both the coupling strength and the effective inertia of the emitted state.  This enforces a color equivalence principle that renders the temperature universal to the background, $T_c \sim 1/Q$.  This universality arises because the relevant worldline instanton corresponds to a topological winding of the probe around the color singularity of the source. In the massless limit the suppression reduces to a pure residue, directly dual to the horizon pole that fixes the Hawking temperature in gravity.

Furthermore, by incorporating the depletion of the finite source, we derive a quadratic backreaction correction, $X \sim \alpha \lambda - \beta \lambda^2$.  This term maps precisely to the Parikh-Wilczek correction for Hawking radiation~\cite{Parikh:1999mf}, confirming that ``Einstein requires Yang-Mills'' not just for the classical metric, but for the thermodynamics of the horizon itself.

The paper is organized as follows.  In \Cref{sec:worldline}, we briefly review the worldline formalism for vacuum persistence.  In \Cref{sec:nonabelian}, we derive the instanton action for the non-abelian root, establishing the linearity of the exponent and the emergence of the color temperature.  \Cref{sec:backreaction} extends this analysis to include dynamical screening, deriving the quadratic backreaction correction and its representation-theoretic origin.  Finally, in \Cref{sec:abelian_contrast}, we contrast these results with the abelian Schwinger effect to highlight the unique features of the non-linear double copy.  We conclude with future directions and phenomenological applications in \Cref{sec:conclusion}. 

\section{Vacuum Persistence and the Worldline Formalism}
\label{sec:worldline}

The worldline representation of one-loop effective actions provides a unifying description of scattering amplitudes, vacuum persistence, and semiclassical decay~\cite{Bern:1991aq, Strassler:1992zr, Schubert:2001he}. In the present work we focus on closed Euclidean worldlines, which isolate the imaginary part of the effective action and describe vacuum decay via worldline instantons.

A convenient starting point is the vacuum-to-vacuum transition amplitude
\begin{equation}
\mathcal{Z} = \langle 0_{\rm out} | 0_{\rm in} \rangle
 = e^{  \frac{i}{\hbar} \Gamma[A]},
\end{equation}
where $\Gamma[A]$ is the one-loop effective action in the external field $A_\mu$.
Throughout this paper we work in units where  $\hbar=1$; restoring $\hbar$ makes explicit that the decay rates are non-perturbative in the semiclassical limit $\hbar\to0$.

The imaginary part of the effective action controls vacuum decay:
\begin{equation}
P_{\text{vac}} = |\mathcal{Z}|^{2}
 = \exp\left(- 2 \Im \Gamma[A]\right).
\end{equation}
The probability of pair production per unit volume and time interval, or pair production rate, is then:
\begin{equation}
\mathcal{V} P_{\text{pair}} = 1-P_{\text{vac}} \approx 2 \Im \Gamma[A]\,.
\end{equation}
Here $\mathcal{V}$ is the  spacetime volume which we will set to unity for the rest of the manuscript.
The approximation holds when the field is weak and $\Im \Gamma[A]$ is small, which is the regime of valid instanton methods we will employ.  Our $P_{\text{pair}}$ is occasionally referred to as  $\Gamma$ in the literature~\cite{Parikh:1999mf,Parikh:2004ih}, where we will reserve $\Gamma$ to refer to the effective action.

For a scalar field in an external background gauge field, working in the mostly-plus convention used throughout this paper,
\begin{equation}
\mathcal{L} = -(D_\mu \phi)^\dagger (D^\mu \phi) - m^2 \phi^\dagger \phi,
\qquad D_\mu = \partial_\mu - i g A_\mu,
\end{equation}
the one-loop effective action can be written 
\begin{equation}
\Gamma[A] = i  \Tr\ln(-D^2 +m^2 -i \epsilon)\,,
\end{equation}
where the trace is over spacetime and any internal indices, and we have used 
the fact that $\ln(D^2 - m^2+ i \epsilon) = \ln(-D^2 + m^2-i \epsilon) + \ln(-1)$.   
We keep the $i \epsilon$ to allow for applying Schwinger's trick in Minkowski space before Wick rotating.

Introducing the proper-time representation of the logarithm,
\begin{equation}
\ln(X-i\epsilon)
   = -\int_0^\infty \frac{dT}{T} e^{-iT(X-i\epsilon)},
\end{equation}
allows us to rewrite the effective action as
\begin{equation}
\Gamma[A]
 = - i  \int_0^\infty \frac{dT}{T}   \Tr \left(e^{-iT(-D^2+m^2-i\epsilon)}\right),
\label{eq:propertime}
\end{equation}
which is the starting point for the Schwinger proper-time and
heat-kernel/worldline representations of the one-loop effective action.

Inserting a complete set of position eigenstates turns the trace into a
quantum-mechanical path integral over closed worldlines $x^\mu(\tau)$.  Upon
Wick-rotating to Euclidean proper time and spacetime,
\begin{equation}
t \to -i t_E, \qquad T \to -i T_E,
\end{equation}
the Euclidean effective action for an abelian theory with $T_E$ relabelled as $T$ for notational
simplicity, becomes
\begin{equation}
\Gamma_E[A]
=  \int_0^\infty \frac{dT}{T} 
 \int \mathcal{D}x 
 \exp \left[  -\int_0^T d\tau 
    \left( \frac{\dot{x}^2}{4} - i g \dot{x}^\mu A_\mu(x) + m^2  \right)  
    \right],
\label{eq:worldline_action}
\end{equation}
under periodic boundary conditions $x^\mu(0)=x^\mu(T)$.  We followed implicitly the phase convention of $\Gamma_E=i \Gamma_M$.  We will defer a discussion of path-ordering in the non-abelian theory to the next section.

The imaginary part of the Minkowski effective action $\Gamma[A]$ arises from complex saddle points
of this Euclidean worldline path integral.  These saddles are closed trajectories in
Euclidean time --- worldline instantons --- which describe the semiclassical
decay of the background field to on-shell physical states~\cite{Affleck:1981bma,Dunne:2005sx}.
Evaluating the Euclidean action on such saddles yields an imaginary contribution to
$\Gamma[A]$, which controls vacuum decay probabilities.  

In the semiclassical limit, as we will now see, this contribution can be extracted from the Euclidean Hamilton-Jacobi action evaluated along the corresponding instanton trajectories.

\subsection{Reduction to a One-Dimensional Problem}

In a static, spherically symmetric background, the classical equations of motion reduce to radial motion in what amounts to an effective potential $V_{\text{eff}}$.  The vacuum decay rate is thereby determined by the action of the dominant worldline instanton trajectory $\bar{r}(\tau)$,
\begin{equation}
\text{Im} \Gamma \approx \mathcal{N} \exp\left( - S_{\text{WL}}[\bar{r}] \right), 
\qquad 
S_{\text{WL}}[r] = \int d\tau \left(\frac{\dot{r}^2}{4} + V_{\text{eff}}(r)\right).
\end{equation}
Here $\mathcal{N}$ represents the one-loop functional determinant of fluctuations around the saddle, which we suppress in the following ($\Im \Gamma[A] \sim e^{-X}$) to focus on the exponential dependence. The classical instanton solution,  $\bar{r}(\tau)$,  extremizes the Euclidean worldline action. 

We can find the path that extremizes the action by enforcing the the vanishing of the inverse scalar propagator dressed by the background field. In Euclidean space the on-shell inverse propagator can be written in Hamilton-Jacobi form as,
\begin{equation}
(p+ i g A)^2=m^2\, ,
\label{eq:on_shell}
\end{equation}
with $p_\mu$ being the canonical momentum.  This is the condition that the classical trajectory lies on the support of the background-dressed scalar propagator.    In this context, $V_{\rm eff}(r)$ can be viewed as a convenient rewriting of the radial on-shell
constraint.   An efficient method of solving \Cref{eq:on_shell} is to introduce the Hamilton-Jacobi (H-J) principal function $S$ with separable ansatz, so that $p_\mu = \partial_\mu S$.

On the leading saddle responsible for $\Im\Gamma[A]$, the Euclidean worldline action reduces to a geometric phase-space area, i.e. the WKB phase integral.  The imaginary part of the effective action is therefore controlled by the on-shell radial canonical momentum $p_r$ evaluated on the dominant worldline saddle,
\begin{equation}
\ln( \Im \Gamma[A])\propto -S_{\rm WL}[\bar r]=-X = -\oint p_r \, dr.
\label{eq:generic_instanton}
\end{equation}
Here the closed phase-space contour $\oint p_r dr$ encodes the analytic structure of the background-dressed propagator and provides a compact representation of the semiclassical contribution to $\Im\Gamma[A]$~\cite{Akhmedov:2006pg,Ilderton:2015lsa,Ilderton:2015qda}.  
In the massive case (and absent other relevant poles) this contour integral reduces to the familiar WKB barrier integral between classical turning points, determined by the zeros of $V_{\rm eff}(r)$.

From an amplitudes perspective this is a semiclassical realization of unitarity. The contour integral enforces the on-shell condition for the background-field dressed probe propagator. 
The imaginary part of the effective action arises from effectively cutting the probe loop.

For the Coulomb-like backgrounds of interest in this paper, the massless limit removes 
all classical turning points, and the canonical momentum becomes dominated by the
gauge (or gravitational) potential.  The leading contribution is then
determined entirely by the topology of this potential: in Yang-Mills the
relevant singularity is the simple pole at the physical source $r=0$, while
in gravity it is the simple pole at the event horizon.  In this limit, the
instanton contribution reduces to a residue.  

The massive and massless regimes are thus unified within the same one-dimensional
worldline action, differing only in whether the on-shell contour encloses a finite barrier or a topological singularity.

\subsection{Review: Schwinger and Hawking as Vacuum Decay}

Before turning to the non-abelian Yang-Mills root, it is useful to briefly recall
how the standard worldline instanton structure arises in two familiar settings:
Schwinger pair production in a constant electric field and Hawking emission near
a black hole horizon.  In both cases the problem reduces to a one-dimensional
Euclidean worldline problem with an effective one-dimensional potential, so that the
semiclassical exponent is fixed entirely by the geometry of the on-shell contour.

\subsubsection{Schwinger effect}
For a charged particle, charge $q$ and mass $m$, in a constant electric field $E$, the potential energy is
$U(x) = -qEx$, and the effective worldline action reduces to a bounce between 
classical turning points $x_\pm = \pm m/(qE)$.  Evaluating the closed WKB action
along the Euclidean trajectory yields the familiar Schwinger exponent
$X_S = \pi m^2/(qE)$, so that the vacuum decay rate scales as
$P_S \sim \exp(-X_S)$.
This is the textbook realization of pair creation as tunneling through a linearly varying potential barrier.

\subsubsection{Hawking radiation}
\label{sec:geometric_hawking}

Before analyzing the Yang-Mills root, it is especially instructive to recall venerable derivations of thermality in gravity.   Here we see Hawking thermality arise as a topological constraint~\cite{Gibbons:1976ue, Hartle:1976tp}.

In Euclidean signature ($t_E = i t$), the Schwarzschild metric near the horizon ($r \to 2 \Gn M$) takes the form of Rindler space.  Defining the proper radial distance $\rho = \int \sqrt{g_{rr}} dr \approx \sqrt{8 \Gn M(r-2 \Gn M)}$, the near-horizon line element becomes:
\begin{equation}
ds^2 \approx \rho^2 \left( \frac{dt_E}{4 \Gn M} \right)^2 + d\rho^2 + r_H^2 d\Omega^2.
\end{equation}
The geometry describes a plane in polar coordinates $(\rho, \Theta)$, with  $\Theta = t_E/4 \Gn M$. Regularity at the origin requires periodicity of the angular coordinate, fixing the circumference of Euclidean time to be  $8\pi \Gn M$.

In the worldline formalism, the vacuum decay rate is determined by the action of a probe winding\footnote{An approach closely related to the tunneling formulation of Parikh and Wilczek~\cite{Parikh:1999mf} and its coordinate-invariant extension for general horizons~\cite{Angheben:2005rm}.} around this topological defect~\cite{Akhmedov:2006pg}.  For a stationary particle of energy $\omega$, the Euclidean worldline action is simply the energy integrated over the temporal circumference:
\begin{equation}
X_H = \omega \oint dt_E = \omega (8\pi  \Gn M).
\end{equation}
The resulting semiclassical suppression factor, $P_H \sim e^{-X_H}$, yields the Boltzmann distribution with temperature $T_H = 1/8\pi \Gn M$.

This derivation does not rely on a potential barrier or a mechanical bounce.  The thermality is purely topological: the mass $M$ creates a defect in the manifold that can only be regularized by a periodic identification of the dual coordinate (Euclidean time).  The question for the double copy is whether the gauge theory root exhibits a dual topological structure in its own characteristic variables.

These two examples demonstrate that particle production in these types of background fields
shares a common semiclassical origin: the rate is governed by the geometry of a
   worldline contour arising in a one-dimensional reduction of the
worldline action.  The structural similarity between the exponents is already
suggestive, yet in the purely abelian setting the correspondence remains
formal.  QED lacks the internal charge structure required for a genuine double copy
to Einstein gravity, and the agreement at leading order is therefore only an abelian 
shadow of a deeper mechanism.  To expose the origin of thermality
and its relation to the double copy, we must turn to the Yang-Mills root background.


\subsection{Yang-Mills Backgrounds and the Worldline}
Internal (color) degrees of freedom have been incorporated in the worldline
formalism using auxiliary Grassmann variables~\cite{Mueller:2019gjj}, yielding
classical color charges in the semiclassical limit and reproducing Wong-type~\cite{Wong:1970fu}
equations of motion.  

An isolated Yang-Mills point source effectively abelianizes its classical field~\cite{Sikivie:1978sa}.  In this work we will consider the abelianized case where the background gauge field admits a fixed color orientation. The worldline under consideration is that of a scalar probe propagating in
this external field. As we will discuss, the only nontrivial effect is encoded in the analytic structure of the trajectory, where winding in the complexified radial plane induces a phase rather than a rotation in color space.

Vacuum analyses of the Yang-Mills effective action (see e.g.~Ref.~\cite{Savvidy:2019grj} and references therein) emphasize that long-wavelength observables depend only on the invariant field strength rather than microscopic structure. Our instanton saddle exploits a similar insensitivity in the presence of a source-supported background.

We now turn to developing the precise realization of this simplification, and 
its physical justification in terms of macroscopic color sources.


\section{Non-abelian Origin of Thermality}
\label{sec:nonabelian}
We now apply the worldline formalism to understand, in the context of a semiclassical worldline process, the color thermality calculated via dynamical scattering amplitudes in Ref.~\cite{Carrasco:2025bgu}, and to extend it to include non-perturbative backreaction.
We analyze the non-abelian Yang-Mills root of a Schwarzschild background.  
Consistent with the scattering analysis, we consider a long-range, Coulomb-like gauge field,
\begin{equation}
 A^a_\mu  = c^a  \frac{Q}{r} k_\mu ,
\end{equation}
generated by a macroscopic composite source of coherent color charge $Q$.   The gauge-invariant quadratic Casimir of the macroscopic source $C_2(R)=\hat{Q}^a \hat{Q}^a$ provides a natural identification for the magnitude of the source color vector,
\begin{equation}
 Q \sim \sqrt{C_2(R)} \,.
\end{equation}
We hold $c^a$ in a fixed color direction as an exact classical solution, effectively abelianizing the gauge field in the root-Kerr-Schild gauge where $k^\mu = (1,1)_{t,r}$.  Note that the vector is null, $k^2=0$, which means we do not have to worry about any contact terms that involve the background interacting with itself, such as the four-point contact in minimal coupling normally required for gauge invariance.

This setup imposes three immediate constraints on the theory:
\begin{enumerate}
 \item The gauge symmetry must remain unbroken to support massless vectors and long-range fields.
 \item The theory must be non-confining on the relevant length scales of the worldline dynamics.
 \item The source representation must therefore be large enough to support a dense hierarchy of Casimirs, $Q \gg 1$, allowing for a quasi-continuous spectrum of backreacted states.
\end{enumerate}

These conditions generally exclude the Standard Model vacuum as SU(3)$_c$ confines at $\Lambda_{\rm QCD}^{-1}$, forcing asymptotic sources to be singlets, and SU(2)$_L$ is Higgsed, screening the long-range interaction.  That said, relevant regimes are natural in conformal (or effectively conformal) settings.   Indeed, $\mathcal{N}=4$ super Yang-Mills has long served as a canonical laboratory for the double copy\footnote{Note the duality between color and kinematics does not require non-confining behavior, e.g.~QCD with massive quarks in the fundamental with $N_f=3$ is color dual which has been noted at both tree and loop levels~\cite{Johansson:2014zca,Johansson:2015oia,delaCruz:2015dpa,Johansson:2017bfl,Kalin:2018thp,Duhr:2019ywc,Kalin:2019vjc,Mogull:2020sak,Carrasco:2023vjg}.}.

 Strict validity of the semiclassical saddle-point approximation requires the action to be large, $X_c \gg 1$.  Our description is rigorous for macroscopic, coherent color sources where $g Q |\lambda| \gg 1$, dual to the requirement $\Gn M \omega \gg 1$ for semi-classical black holes.  

Against this fixed-orientation Yang-Mills background we look at the instanton associated with a scalar probe in the adjoint.  For a generic Yang-Mills background the Euclidean action involves a path-ordered exponential,
\begin{equation*}
\int \mathcal D x
 \exp\left[ -\int_0^T d\tau \left( \frac{\dot x^2}{4} + m^2 \right) \right] \,\,
\mathcal P \exp\left( + i g \int_0^T d\tau \dot x^\mu A_\mu^a(x) T^a \right).
\end{equation*}
Our background field is quite special with fixed color-orientation $c^a$. At this point the Wilson loop becomes, 
\begin{equation*}
\exp\left( + i g \int_0^T d\tau \dot x^\mu  A_\mu(x) (c^aT^a) \right).
\end{equation*}
We chose the basis of the probe field to diagonalize the operator $c^a T^a$ so the scalar particle contributes its eigenvalue $\lambda$.  
Our Euclidean action thereby abelianizes in this background to
\begin{equation}
\Gamma_E[A]
=  \int_0^\infty \frac{dT}{T}  \int \mathcal{D}x  
 \exp\left[ -\int_0^T d\tau \left( \frac{\dot{x}^2}{4} - 
 i g Q \lambda   \dot{x}^\mu \frac{k_\mu}{r} + m^2 \right) \right].
\label{eq:ym_worldline_action}
\end{equation}

We define the effective abelian charge $\alpha$ via,
\begin{equation}
i g Q \lambda\frac{1}{r} k_\mu \phi \equiv i \frac{\alpha}{r} k_\mu \phi \, .
\end{equation}
From this point onward we work in the diagonalized sector and denote by
$\mathcal{A}_\mu$ the effective abelian background felt by the scalar probe. 
So, e.g.~$L_E = \dot{x}^2/4 - i \mathcal{A}\cdot \dot{x} + m^2$.

We will consider both massive and massless scalar probes in the adjoint.  As discussed above, a worldline instanton in a background field is often understood as a tunneling process through a classically forbidden region: the Euclidean action receives an imaginary contribution from integrating the radial momentum between turning points where the radial velocity vanishes.  This picture applies directly to the massive probe or attractive eigenvalues ($\alpha<0$) and reproduces the familiar bounce instantons of Schwinger pair production.

In the massless limit, however, the structure simplifies dramatically.  The canonical momentum is dominated by the gauge potential.  We argue that the leading contribution is then determined not by the detailed shape of an effective potential, but by a topological winding of the worldline around the pole in the gauge background at the origin.  We will show that the semiclassical exponent becomes a pure residue, 
\begin{equation}
|X_c| = 4 \pi g Q |\lambda| \, , 
\end{equation}
independent of trajectory details or radial dynamics.  

\subsection{Color Instantons and Vacuum Decay}

Our goal is to resolve the exponential behavior of the decay amplitude $\text{Im} \Gamma[A] \sim e^{-S_E}$. In the WKB limit, the Euclidean action $S_E$ (\Cref{eq:worldline_action}) satisfies the on-shell condition.  Using the covariant derivative in terms of our effective abelianized field, $D_\mu= \partial_\mu - i \mathcal{A}_\mu$, this is simply,
\begin{equation}
g^{\mu\nu}_E (\partial_\mu S + i \mathcal{A}_\mu) (\partial_\nu S + i \mathcal{A}_\nu) - m^2=0\,.
\label{eq:oneLoopCut}
\end{equation}

We assume the separable ansatz for the H-J principal function,
\begin{equation}
S = E \tau + W(r) \, ,
\end{equation}
so that the momentum components are energy $E$ and radial momentum $p_r$,
\begin{equation}
\partial_\tau S=E, \qquad \partial_r S = (p_r \equiv W'(r)).
\end{equation}

Substituting the Euclidean metric and background field components $\mathcal{A}_\tau = -i \alpha/r$, and $\mathcal{A}_r = \alpha/r$, we can solve \Cref{eq:oneLoopCut} for the radial momentum,
\begin{equation}
 p_r =- i \frac{\alpha}{r} \pm \sqrt{m^2 - (\frac{\alpha}{r} + E)^2} \, .
\end{equation}
We see both a real and imaginary component to the radial momentum.

Now let us consider them both in turn, finding that the real component yields a classically forbidden region only in the massive case, and the imaginary component yields the universal color thermality discovered via S-matrix methods in Ref.~\cite{Carrasco:2025bgu}.

\subsection{The Massive Barrier}
\label{sec:massive}
The massive aspect of the suppression $\Gamma \sim e^{-S_{\text{real}}}$ is governed by the integral of the real part of $p_r$ between the classical turning points:
\begin{equation}
    S_{\text{barrier}} = \oint \Re p_r dr = 2 \int_{r_{in}}^{r_{out}} \sqrt{m^2 - \left(\frac{\alpha}{r} + E \right)^2}  dr.
\end{equation}
The turning points occur where  the square root vanishes and for $r\geq0$.  Real turning points require $\alpha<0$, so we write $\alpha = -|\alpha|$, finding
\begin{equation}
 E -    \frac{|\alpha|}{r}  = \pm m \implies r_{in} = \frac{|\alpha|}{E+m}, \quad r_{out} = \frac{|\alpha|}{E-m}.
\end{equation}
Clearly there are no turning points when $m \to 0$.  Also note, the turning points are only physical $\iff \lambda <0$, i.e.~when $\lambda$ is attractive to $Q$ (taken positive).   Repulsive $\lambda$ experience no massive barrier. The barrier integral then yields the massive contribution to the WKB suppression factor,
\begin{equation}
 S_{\text{barrier}} =2  \pi \, |\alpha | \left(\frac{E}{\sqrt{E^2-m^2}} - 1\right)\,.
 \end{equation}
The integral across the forbidden region provides the suppression due to the mass gap.  Three features are particularly noteworthy: 
\begin{enumerate}
\item The barrier contribution diverges at threshold ($m=E$).
\item The barrier contribution vanishes at high energy ($E\gg m$).
\item The barrier contributes only for $\alpha<0$ and $Q>0$.
\end{enumerate}

\subsection{Massless Limit, the Aharonov-Bohm Phase, and Universal Thermality}
\label{sec:massless}

In the limit $m\to0$, we see that the spatial  barrier disappears.  The surviving imaginary part of $p_r$ still contributes to $\oint p_r dr$ through a winding around the singularity that controls the universal thermal weight. In essence we can understand this as a non-abelian Aharonov-Bohm type effect. 

It is instructive to consider the derivation of Hawking radiation  in Kerr-Schild coordinates.
In Kerr-Schild-type coordinates for the gravitational case, the background remains smooth, but the canonical momentum of a probe develops a singularity from crossing the light-cone at the event horizon.  The gravitational computation is summarized in \Cref{sec:gr_kerr_schild}, where the same contour prescription produces
$P^{\rm GR}_{\rm pair}\sim e^{-8\pi G_N M\omega}$.
The role of the horizon pole in gravity is thus directly mirrored by the color pole in the root theory.

 In our case, for a massless scalar in the dual gauge theory root-Schwarzschild background, the radial canonical momentum is:
\begin{equation}
 p_r =- i \frac{\alpha}{r} \pm i \left(\frac{\alpha}{r} + E \right).
 \end{equation}
 
 We find a regular $(+)$ branch where $p_r = i E$.  But critically note that the other branch $(-)$ carries a pole at $r=0$,
 \begin{equation}
 p_r = -2 i \frac{\alpha}{r} - i E.
 \end{equation}
It is precisely around this pole that we must define a contour in the complexified $r$-plane.

We evaluate the winding, picking up the pole,
\begin{equation}
    S_{\text{winding}} = \oint p_r  dr =  \oint \left( -2 i \frac{\alpha}{r} \right) dr 
    = \pm 4 \pi \alpha \,,
\end{equation}
where the sign is selected by the positive Feynman $i\epsilon$ prescription to ensure a positive Euclidean action for whatever sign $\lambda$ carries  and hence an exponentially suppressed decay rate.

The magnitude of the action accumulated during this crossing is
\begin{equation}
 |S_{\text{winding}}| = 4 \pi |\alpha| \, .
\end{equation}
The corresponding emission probability rate is therefore dominated by
\begin{equation}
P_{\text{pair}} \approx    \exp(-|S_{\text{winding}}|)  =  \exp(- \frac{|\lambda|}{T_c} ) \,, \qquad
  T_c \equiv \frac{1}{4 \pi  g Q} .
  \label{eq:color_temperature}
\end{equation}

It is important to note that this contribution due to winding is present even in the case of massive scalar fields, but the rate will be further suppressed in that case due to $S_{\text{barrier}}$.  

 This result satisfies all physical limits:
\begin{itemize}
    \item \textbf{Massless Limit ($m \to 0$):} The barrier contribution becomes unity ($\exp(0)=1$). We recover the pure thermal spectrum derived in the eikonal scattering paper~\cite{Carrasco:2025bgu}.
    \item \textbf{Threshold Limit ($E \to m$):} The exponential argument diverges ($\exp(-\infty) = 0$). Emission shuts off at the kinematic threshold.
    \item \textbf{High Eigenvalue Limit ($\lambda \to \infty$):} Emission of large color charge eigenvalues are suppressed by both the thermal factor and any mass factor.
\end{itemize}
This confirms that  the color thermality is a robust feature of the non-abelian root, modified --- but not destroyed --- by the mass of the probe.

Although the process is pair production, we can build intuition with a semiclassical story: Only particles repelled by the background abelianized charge appear as asymptotic states. The oppositely charged partner is attracted to the source and falls inward. The semiclassical exponent therefore governs the probability to produce an escaping particle. The massive  barrier is encountered only by the attracted particle. Because pair production is a correlated process, the semiclassical rate for emitting an asymptotic particle includes a single barrier suppression factor.

The color horizon of the coherent source is therefore not a spatial surface but a topological boundary defined by the enclosed color flux. Once the leading instanton is understood as winding around the relevant analytic pole, the backreaction analysis of \Cref{sec:backreaction} becomes physically intuitive.   Pair production creates charges $\pm\lambda$ simultaneously.  As a result, subsequent worldline windings enclose a reduced effective charge $Q' = Q - |\lambda|$, and therefore accumulate a smaller winding phase.   Integrating this depletion yields (upon double copy) the quadratic Parikh-Wilczek correction, encoding the finite capacity of the source to radiate.

\subsection{Relation to the S-matrix derivation of Root Thermality}
In the companion analysis of Ref.~\cite{Carrasco:2025bgu}, the same color-thermal spectrum was derived directly from the eikonal S-matrix in real time, treating radiation in the soft limit of the collapsing shell background.  There, the radiation spectrum was extracted from the overlap of a probe state with its asymptotic out-state.  In that framework, the thermal distribution arose from the Fourier transform of a logarithmic eikonal phase generated by the long-range potential.  In particular, it was shown that the dynamical prefactor controlling the emitted spectrum is thermal in the color eigenvalue $\lambda$ and that the semiclassical evolution of the spectrum is governed by the available density of color states.

The worldline instanton result provides a non-perturbative confirmation of this structure.  The two approaches operate in distinct regimes: the S-matrix analysis performs a systematic soft resummation in real time, while the present treatment computes vacuum decay through semiclassical decay in Euclidean time.  Their agreement at leading order and through the $\mathcal{O}(\lambda^2)$ correction establishes the robustness of color thermality and demonstrates that its origin is non-perturbative rather than an artifact of the eikonal limit.  Moreover, the derivation here exposes explicitly the non-abelian character of the effect through the $\lambda$-dependence of the exponent, rather than through soft factors in momentum space.

It is worth noting that the single-instanton saddle point derives the Boltzmann
suppression factor,
\begin{equation}
\frac{dN}{d\lambda} \sim e^{-|\lambda|/T_c},
\end{equation}
rather than the full Planck-like distribution,
\begin{equation}
\frac{dN}{d\lambda}  \propto \frac{|\lambda|/T_c}{\sinh(|\lambda|/T_c)} \times [\text{phase-space factor}],
\label{eq:inclusive_rate}
\end{equation}
obtained in the scattering amplitude analysis of
Ref.~\cite{Carrasco:2025bgu}.
This is natural.
The quantity $\beta/\sinh\beta$ appearing in the S-matrix should be understood as
the universal periodic spectral function associated with Gaussian
fluctuations on a circle.
The worldline instanton computes the vacuum persistence amplitude for a single
periodic loop, fixing the monodromy parameter $\beta$ and hence the leading
Boltzmann factor.

The full Planckian form arises automatically upon summing over multi-instanton
sectors, corresponding to multiple windings of the worldline around the
Euclidean time circle~\cite{Affleck:1981bma,Dunne:2005sx}.
Equivalently,
\begin{equation}
\frac{\beta}{\sinh\beta}
=
e^{-\beta}
\Bigl(2\beta\sum_{n=0}^{\infty} e^{-2n\beta}\Bigr),
\label{eq:resummed_Thermality}
\end{equation}
where the single-instanton saddle supplies the exponential suppression while the
sum over windings reconstructs the full thermal spectrum.
Our result therefore fixes the dynamical origin of the temperature $T_c$ --- the
monodromy parameter controlling the exponential suppression --- while the detailed
infrared spectral shape ($\lambda\ll T_c$) is dictated by the quantum statistics
of the probe.

\section{Backreaction and the Quadratic Correction}
\label{sec:backreaction}

The analysis in \Cref{sec:nonabelian} captures the leading semiclassical structure of vacuum decay in the probe limit, where the background field is treated as an infinite reservoir of charge.  In this limit, the source $Q$ is fixed, and the resulting spectrum is strictly thermal.

To extend beyond this approximation, we must account for the finite nature of the source.  As an on-shell particle is radiated, conservation of charge requires the source to deplete.  In the gravitational context, Parikh and Wilczek~\cite{Parikh:1999mf} demonstrated that enforcing such conservation laws dynamically modifies the spectral suppression, introducing non-thermal corrections to the exponent.  

It is crucial to distinguish the effect in this section from standard perturbative loop corrections.  While one-loop vacuum polarization typically renormalizes the effective coupling or modifies the prefactor of the decay rate, the backreaction considered here arises from the change in the background state itself, enforced by conservation of color charge during radiative decay.  This modifies the geometry of the instanton solution, producing corrections to the exponential action that depend non-linearly on the emitted charge.  This should be understood as modifying the $\beta$ in \Cref{eq:resummed_Thermality}.

\subsection{Dynamical Screening}
We quantify this effect by treating the background charge as a dynamical variable.  As an emitted on-shell  particle with color eigenvalue $\lambda$ radiates to spatial infinity, its attractive partner screens the background source.  Conservation of color charge (and indeed Gauss's law) implies the instantaneous substitution
\begin{equation}
Q  \longrightarrow  Q(\lambda') = Q - \lambda',
\label{eq:screening_shift}
\end{equation}
where $\lambda'$ represents the charge that has already escaped.  The full instanton action is obtained by integrating the differential cost over the build-up of the emitted charge:
\begin{equation}
X_{\mathrm{full}}(\lambda) = \int_0^{|\lambda|} d\lambda'  \frac{\partial X_c}{\partial \lambda'} \Big|_{Q \to Q(\lambda')}.
\end{equation}
Using the linear leading-order result $X_c \approx  4\pi g Q \lambda'$, the integrand is simply $4\pi g (Q - \lambda')$.  Performing the integration yields
\begin{equation}
X_{\mathrm{full}}(\lambda) 
= 4 \pi g  |\lambda|  \left (Q -    \frac{ |\lambda|}{2}\right)\, .
\label{eq:quadratic_correction}
\end{equation}

The corresponding pair production rate acquires a quadratic correction in the exponential (valid in the regime $|\lambda|\ll Q$),
\begin{equation}
P_{\text{pair}} \sim \exp\left[-X_{\mathrm{full}}(\lambda)\right]
 \propto  
\exp\left[-4 \pi g  |\lambda|  (Q -     |\lambda|/2)\right]\,.
\label{eq:correction_decay}
\end{equation} 

This result possesses  two significant features.  First, the quadratic term represents a deviation from pure thermality, encoding the type of correlations between emitted quanta required for a unitary S-matrix.  Second, the sign of the correction  indicates an enhancement of the decay rate relative to the Boltzmann suppression.  

This behavior is directly dual to the Parikh-Wilczek correction in gravity, 
\begin{equation}
P^{\text{GR}}_{\text{pair}} \sim \exp \left[ -8 \pi \Gn  \omega(M  - \omega/2) \right ]\,.
\end{equation} 
Just as the reduction of black hole mass reduces the gravitational suppression, the dynamical screening of the background charge increases the production rate.  The correspondence is precise: the non-linear backreaction of the gauge theory is the double copy dual of the backreaction of the geometry. As in gravity this backreaction-induced correction ensures that successive emissions are not independent.  We can recognize this as a semiclassical imprint of the correlations required by a unitary $S$-matrix.

\subsection{Representation-Theoretic Interpretation}
\label{sec:rep_theory}

The quadratic correction derived from dynamical screening admits a sharp interpretation in terms of the representation theory of the color source.  Rather than treating the backreaction as a separate dynamical effect, we view the entire background decay process as a transition between representations of the background source.

Consider the macroscopic source to be in a representation $R$ with quadratic Casimir $C_2(R) \sim Q^2$. The emission of a particle with color eigenvalue $\lambda$ induces a transition to a new representation $R'$,
\begin{equation}
 R  \longrightarrow  R' \otimes r_\lambda,
\end{equation}
where $r_\lambda$ is the representation of the emitted quantum.  Conservation of color charge implies that the effective charge of the source is reduced, $Q' = Q - \lambda$.  The associated change in the Casimir governs the change in the entropy of the configuration:
\begin{equation}
 \Delta C_2 = C_2(R) - C_2(R') \propto Q^2 - (Q-\lambda)^2.
\end{equation}
Expanding this difference yields two distinct contributions:
\begin{equation}
 \Delta C_2 \propto \underbrace{2Q\lambda}_{\text{Thermal}}  -  \underbrace{\lambda^2}_{\text{Correction}}.
\end{equation}

The leading linear term corresponds to the Boltzmann suppression $\exp(-|\lambda|/T_c)$ identified in \Cref{eq:color_temperature}, confirming that the temperature is simply the derivative of the Casimir density of states.  The subleading term provides the quadratic correction:
\begin{equation}
 \delta X_c \propto - g \lambda^2.
\end{equation}
 The Parikh-Wilczek derivation relies on the fact that the emission probability is determined by the change in the black hole entropy, which equals the change in the horizon area: $\Delta S_{\text{BH}} \propto \Delta (M^2)$.  In the double copy, energy is the charge of gravity. Consequently, in the rest frame of the black hole, the squared mass $M^2$ is the gravitational dual of the quadratic Casimir.  

Our result demonstrates that this mapping holds dynamically: the gauge theory production rate is governed by $\Delta C_2$ exactly as the gravitational rate is governed by $\Delta M^2$.  Thus, the quadratic deviation from thermality is not an isolated feature of black holes, but the precise dual of the finite-rank change in the color representation.

\subsection{The Double Copy Dictionary}

We comment here on the  refinement that backreaction offers to the double copy mapping between the non-abelian root and gravity structures. The linear dependence on $\lambda$, due to WKB winding, reflects the universal $1/r$ singularity. The quadratic correction reflects the finite capacity of the source.

Applying the dictionary:
\begin{center}
\begin{tabular}{c c c}
    \textbf{Yang-Mills} & $\longleftrightarrow$ & \textbf{Gravity} \\ \hline
    Color Charge $Q$ & $\longleftrightarrow$ & Mass $M$ \\
    Eigenvalue $\lambda$ & $\longleftrightarrow$ & Energy $\omega$ \\
    Quadratic Casimir $C_2\sim Q^2$ & $\longleftrightarrow$ & Horizon Area $A \sim M^2$ \\
    $4\pi g |\lambda| (Q - \tfrac{1}{2}|\lambda|)$ & $\longleftrightarrow$ & $8\pi \Gn \omega  (M - \tfrac{1}{2}\omega)$
\end{tabular}
\end{center}

The linear term maps to the Hawking temperature, while the quadratic term maps exactly to the Parikh-Wilczek correction for finite-mass backreaction. In this way, the non-abelian origin of screening parallels the gravitational origin of backreaction, demonstrating that the structural relationship between the exponents survives beyond leading semiclassical order.  

It is worth commenting on how surprised we should be about any of this. The basic moral is not new.  Since the 1980's  we  have been appreciating in ever sharper contexts the fact that all of Einstein-Hilbert gravity is encoded, in a precise\footnote{See recent double copy reviews~\cite{Bern:2019prr, Borsten:2020bgv,Adamo:2022dcm,  Bern:2022wqg, Bern:2023zkg}, as well as direct operator constructions~\cite{Bern:1999ji,Bern:2010yg,Tolotti:2013caa,Carrasco:2025ymt}.} sense, in the kinematics of Yang-Mills gauge theory.  The surprise is not that gauge dynamics encodes gravitational dynamics.  The surprise is how cleanly the Kerr-Schild Rosetta stone allows us to appreciate features of this correspondence at the level of non-perturbative vacuum physics.  What emerges is a genuinely novel gauge phenomenology bridging effective action and $S$-matrix perspectives.


\section{Shining Light on the Abelian Shadow}
\label{sec:abelian_contrast}

Having established the thermal structure of the non-abelian root and its quadratic backreaction, we are positioned to articulate why the abelian theory --- despite its formal similarities --- cannot serve as a faithful double copy root of Einstein gravity event by event.  The failure occurs at two distinct levels: the universality of the temperature and the nature of the semiclassical backreaction.

\subsection{Universality and the Equivalence Principle}
The structural similarity between the Schwinger and Hawking exponents has long invited comparison, often framed through the lens of the Unruh effect.  In a constant electric field $E$, the Schwinger exponent for a particle of mass $m$ and charge $q$ is $X_S = \pi m^2/(qE)$.  This can be rewritten in terms of the Unruh temperature perceived by the accelerating charge, $T_U = a/2\pi = qE/(2\pi m)$:
\begin{equation}
 X_S = \frac{m}{2 T_U}.
\end{equation}
The rate  $P \sim e^{-X_S}$ thus mimics a Boltzmann suppression.  This works kinematically because the propagator structure of electrodynamics mirrors that of linearized gravity --- much in the same way the  Coulomb force between opposite charges serves as the single copy of the Newtonian gravitational force in regimes valid up to self-interaction.

This abelian reading fails dynamically because it lacks universality. In the Schwinger effect, the effective temperature $T_{\rm eff} \sim T_U$ depends inversely on the inertia of the probe, $T_U \propto 1/m$. So in general electrodynamics, particles of different masses perceive different “temperatures” in the same electric field. This violates the Equivalence Principle required of a dual to Einstein gravity, where the temperature must be a property of the geometry alone, $T_H \propto 1/M$.

This limitation points directly to the necessity of the non-abelian root. In the Yang-Mills instanton (\Cref{sec:nonabelian}), the suppression scales as $X \propto \lambda$, with the color eigenvalue $\lambda$ simultaneously controlling the coupling to the background and the effective inertia of the worldline saddle. Because this ratio is fixed by the algebra, the resulting temperature $T_c \propto (gQ)^{-1}$ depends only on the background source and is independent of the probe mass, $m$.

\subsection{Contrast in ``Backreaction-like'' Effects}
\label{subsec:backreaction_contrast}
The distinction becomes even sharper when mocking up an effect that can superficially resemble an ``abelian root'' of the Parikh-Wilczek backreaction --- but importantly does not involve altering the saddle. This is in contrast to the Yang-Mills backreaction described in \Cref{sec:backreaction}, which fundamentally involved modifying the saddle by depleting the color-source.   We will do so by introducing a Coulombic attraction between the pair in a heuristic\footnote{Here, we are introducing a plausible effective interaction-energy term by hand to probe the structure of the exponent, rather than deriving the full two-loop correction from the worldline path integral.} potential.  This has the effect of reproducing, at least structurally, the contribution of a genuine 2-loop radiative correction~\cite{Affleck:1981bma,Lebedev:1984mei,Huet:2020awq}.  It is important to realize that such radiative corrections are quantum corrections to the same semi-classical background --- quite distinct from the calculations of \Cref{sec:backreaction}.

 In the Schwinger effect, the produced electron and positron do not interact via a non-linear gauge structure, but they do experience a mutual Coulomb attraction which we can include in the potential.  While the Coulomb field does not shift the classical turning points, it can lower the barrier for pair creation.   We model this by adding the Coulomb potential to the external driving field.  The total potential energy for a pair separated by a distance $x$ is:
\begin{equation}
 U(x) = \underbrace{-q E_0 x}_{\text{External Field}} \underbrace{-\frac{1}{2}\frac{q^2}{4\pi x}}_{\text{Mutual Attraction}},
 \label{eq:coul_Correction}
\end{equation}
where we use Gaussian units ($4\pi\epsilon_0=1$), and include a symmetry factor of $\tfrac{1}{2}$ to avoid overcounting the interaction energy.  Treating the Coulomb interaction energy  $U_1(x) = -\frac{1}{2} \frac{q^2}{4 \pi x}$,  as a perturbation to the linear potential $U_0(x) = -qE_0x$, the WKB exponent acquires a first-order correction:
\begin{equation}
 \delta X_S   \approx 2  \int_{x_-}^{x_+} dx  \frac{U_0(x)}{\sqrt{m^2-U_0(x)^2 }} \left(-\frac{q^2}{8\pi x}\right).
\end{equation}
The turning points are determined primarily by the external field, $x_\pm \approx \pm m/(qE_0)$.  Evaluating the integral yields a constant shift:
\begin{equation}
 \delta X_S = -\frac{q^2}{4}.
\end{equation}
The full WKB exponent is thus
\begin{equation}
 X_S(m) = \frac{\pi m^2}{qE_0} - \frac{q^2}{4}.
\end{equation}
We see that the heuristic potential \Cref{eq:coul_Correction} yields a $q^2$ correction and is able to capture at least the leading structure of the two-loop radiative correction~\cite{Affleck:1981bma,Lebedev:1984mei,Huet:2020awq}. 

The contrast with the non-abelian result \Cref{eq:quadratic_correction} is striking.  
\begin{itemize}
\item In the abelian case, rather than a ``backreaction'', the radiative-correction-like $q^2$ term produces a constant shift in the exponent, independent of the inertial mass $m$ (the variable governing the Boltzmann suppression).  This scales the overall pair-production rate without altering the shape of the spectral distribution.  

\item In the Yang-Mills case,  the correction depends quadratically on the color charge eigenvalue $\lambda$ (the variable governing the color thermality).  This modifies the distribution itself, introducing a deviation from pure thermality.  This confirms that the mechanism observed by Parikh-Wilczek in the gravitational case --- where backreaction genuinely alters the spectral shape --- is a specific property of theories with non-linear charge/mass structures (gravity and Yang-Mills), absent in the linear abelian theory.
\end{itemize}

\section{Discussion and Future Directions}
\label{sec:conclusion}

In this work, we have utilized the worldline instanton formalism to realize the thermal nature of radiation from a static, non-abelian source, confirming a prediction made using scattering methods in Ref.~\cite{Carrasco:2025bgu}.  By analyzing the Yang-Mills root of the Kerr-Schild background, we demonstrated that the vacuum decay rate is governed by a temperature $T_c \propto (gQ)^{-1}$.  Furthermore, by incorporating the depletion of the finite source, we derived a quadratic backreaction correction that maps precisely to the Parikh-Wilczek result for Hawking radiation.

The use of one-loop effective actions to reproduce Bogoliubov physics has a long
pedigree including the recent abelian analysis of Ref.~\cite{Ilderton:2025aql}. Our analysis here involves a targeted application of such a 
framework to the Kerr-Schild Yang-Mills root.   This perspective complements several amplitudes-based directions.  
Real-time eikonal resummations extract  apparently thermal spectra in gravitational and non-abelian gauge cases~\cite{Aoude:2024sve, Carrasco:2025bgu} from the $S$-matrix of a
probe in a collapsing background, while absorption-EFT formulations encode
long-distance emission data through Wilson coefficients~\cite{Aoude:2023fdm}.
More ambitiously, a recent proposal has sought to reorganize the entire
Bogoliubov transformation as an on-shell scattering problem for a heavy state~\cite{Aoki:2025ihc}.
Our result isolates the physics that is fixed purely by the
analytic topology of the classical root solution --- independent of any assumed
spectral ansatz or choice of external state --- suggesting a clean separation between
(i) universal monodromy/residue data dictated by conservation laws and (ii)
dynamical information such as greybody factors and short-distance completion.

Both the eikonal S-matrix derivation~\cite{Carrasco:2025bgu} and the current worldline instanton computation are non-perturbative in $1/\hbar$ while remaining  perturbative in the coupling.   Both resum infinitely many interactions with a macroscopic background and produce effects exponentially suppressed in the coupling. Importantly, neither requires new ultraviolet degrees of freedom. The agreement between these approaches reflects the fact that thermality is controlled by the analytic structure of classical solutions rather than by intrinsically quantum-gravitational dynamics.

Our analysis clarifies why the abelian Schwinger effect is an insufficient proxy for gravitational thermality.  While QED shares the propagator structure of linearized gravity, it lacks the non-linear algebraic unity of charge and inertia required to satisfy the Equivalence Principle.  Only in the non-abelian theory does the charge eigenvalue $\lambda$ dictate the effective barrier, enforcing a universal color temperature independent of the probe's mass.

This analysis addresses  long-standing questions regarding the non-perturbative validity of the double copy.  A common critique of the correspondence is that it is merely a perturbative device --- a coincidence of Feynman diagrams that may fail to capture global or topological features of spacetime.  Our results bolster a growing body of work\footnote{This includes the mapping of self-dual instanton topologies~\cite{Berman:2018hwd},
the compatibility of the double copy with S-duality~\cite{Alawadhi:2019urr},  the preservation of topological Aharonov-Bohm phases~\cite{Burger:2021wss}, the embedding of Kerr-Schild double copy in Exceptional Field Theory~\cite{Berman:2020xvs}, as well as the observation that the BFKL equation in QCD~\cite{Kuraev:1977fs,Balitsky:1978ic,Jalilian-Marian:2000pwi}, non-perturbative in energy, double-copies to the Lipatov equation in GR~\cite{Raj:2025hse}.} demonstrating otherwise.   By successfully mapping the non-perturbative topology of the black hole horizon to the winding modes of the gauge theory, we establish that the duality holds at the level of the vacuum structure itself, independent of the perturbative expansion.

Ultimately, this work informs the broader search for the fundamental operator algebras of quantum gravity.  Attempting to decode these algebras directly in the gravitational theory is daunting, as the geometry obscures the algebraic structure.  The single copy offers a strategic advantage: it places the mysterious kinematic algebra alongside the well-understood color algebra.  

Our results demonstrate that the thermodynamic properties of the horizon are not emergent statistical accidents of the geometry, but intrinsic features of the kinematic algebra itself.  The color equivalence principle --- where charge dictates inertia --- is essentially a Ward identity of this underlying structure.  By identifying these features in the gauge theory warm-up, we isolate the algebraic seeds of the horizon, providing a precise target for future work seeking to construct the full operator algebra of the quantum double copy.

Perhaps the most intriguing implication of these results lies in the stability of the source.  Standard intuition suggests that highly charged non-abelian states are inherently unstable.  Our result implies the inverse: because the color temperature scales as $T_c \sim Q^{-1}$, macroscopic sources with large quadratic Casimirs are thermodynamically cold under background field decay.  

So far our entire discussion has been focussed ultimately on recovering  the kinematic aspect of a rate in color eigenvalue: $dN/d\lambda$.  As emphasized in \cite{Carrasco:2025bgu}, an entirely distinct and competing aspect is the color phase space density, cf.~\Cref{eq:inclusive_rate}. To discuss phenomenology of macroscopic coherent sources we must shift this conversation to one about a decay rate in time.

To translate the derived color decay spectrum into a natural time-dependent evaporation process requires additional conditions, entirely presaged by those entering estimates of black hole lifetimes. In particular:
\begin{enumerate}
\item Asymptotic propagation: There must exist color-carrying states that can propagate away from the source and meaningfully count as emitted.  Gravitational dual: asymptotic flatness, allowing Hawking quanta to escape.
\item Weak reabsorption: Emitted quanta should not be efficiently re-captured or scattered back into the source. Gravitational dual: greybody factors admitting escape.
\item Separation of scales: The internal dynamics of the source must remain fast compared to the emission timescale, justifying a semiclassical treatment. Gravitational dual: black holes large compared to the Planck scale.
\item Absence of a confining medium: The environment should not thermalize or trap the radiation before it escapes. Gravitational dual: black holes not embedded in a dense medium or AdS box.
\item Quasiparticle picture: Color eigenstates should behave as well-defined excitations over the relevant scales. Gravitational dual: existence of a particle interpretation for Hawking quanta.
\end{enumerate}
When these conditions hold, stronger suppression in the decay spectrum corresponds directly to rarer events in time. In such a regime, the color-thermal spectrum implies parametrically long-lived macroscopic sources, with an effective evaporation time controlled by the inverse color temperature. As in gravity, more detailed lifetime estimates would depend on model-specific transport dynamics well beyond the universal semiclassical analysis presented here.  

Even so, under such conditions, we can consider the resulting phenomenology. In a confining theory like QCD, macroscopic coherent sources are forbidden by the mass gap.  In a non-confining regime --- such as a hidden sector in an effectively conformal window --- this mechanism permits the existence of coherent, macroscopic color states that are effectively metastable.  These objects would behave like color supernovae: long-lived, coherent configurations that evaporate extremely slowly via thermal radiation in color space, until quite suddenly they pop.

This offers a novel mechanism for beyond the standard model phenomenology.  If a hidden sector admits such large-representation bound states, they could form stable dark matter candidates whose longevity is protected not by a global symmetry, but by the suppression of the Hawking-like color temperature.  

Finally, this mechanism suggests a counter-intuitive frontier for quantum information science.  Typically, macroscopic quantum states are fragile; as a system grows, its coupling to the environment increases, accelerating decoherence.  In stark contrast, our analysis identifies a regime where the opposite is true.  Because the radiation rate $P_{\text{pair}} \sim \exp(-|\lambda|/T_c)$ is exponentially suppressed by the source charge magnitude ($T_c\propto 1/Q$), states in sufficiently large representations of the gauge group become naturally isolated from the vacuum.

This offers a theoretical blueprint for intrinsic error protection.  Rather than engineering a logical qubit through the active error correction of many constituent particles, one could imagine a single macroscopic excitation --- a large-$R$ qudit --- whose coherence is protected by the thermodynamics of the non-abelian force itself.  Just as black holes are theorized to be nature's most robust quantum memories (preserving information until the Page time), their gauge-theoretic duals would represent macroscopic islands of coherence, protected from soft-radiation decoherence not by topology, but by the extreme suppression of the color temperature.

\acknowledgments
We are especially grateful to Gerald Dunne, and Anton Ilderton for valuable discussions, references, and comments on an earlier draft.  Additionally we would like to thank Raju Venugopalan for very instructive conversations about color-glass-condensates, the double copy of BFKL, and important references.  JJMC would like to warmly acknowledge Donal O'Connell and Chris White for related collaboration and clarifying discussions on potential color-source thermodynamics,  Gonzalo Torroba for formative insights on the roots of Hawking radiation and the event horizon, and David Berman for recommending Ref.~\cite{Parikh:1999mf}  towards a double copy analysis during early stages of the pandemic.  This work was supported in part by the DOE under contract DE-SC0015910, and by Northwestern University through the Amplitudes and Insights Group, the Department of Physics and Astronomy, and the Weinberg College of Arts and Sciences.

\appendix 
\crefalias{section}{appendix}

\section{The Gravitational Horizon and Kerr-Schild Regularity}
\label{sec:gr_kerr_schild}

This appendix provides a pedagogical recap of the familiar Hawking computation, cf.~ \Cref{sec:geometric_hawking}, included to highlight the structural parallel with its color-dual Yang-Mills root.

We can now perform a structurally identical analysis to \Cref{sec:nonabelian} for the  gravitational double copy case,  using the Schwarzschild metric expressed in Kerr-Schild form.  In these coordinates, the metric is
\begin{equation}
	g_{\mu\nu} = \eta_{\mu\nu}+\frac{2 \Gn M}{r} k_\mu k_\nu, \qquad k^\mu = (1,\hat{r}),
\end{equation}
where $k^\mu$ is null with respect to both $\eta_{\mu\nu}$ and $g_{\mu\nu}$. The inverse metric in Kerr-Schild form is then simply
\begin{equation}
	 g^{\mu\nu} = \eta^{\mu\nu} - \frac{2 \Gn M}{r} k^\mu k^\nu.
\end{equation}
It is important to realize that here that the metric is well behaved away from the origin.  Even so, every way of writing Schwarzchild leads to a pole in the canonical momentum at the event horizon.  It is this pole --- not any sort of metric coordinate singularity --- that anchors the vacuum decay. 

For a massless probe particle, the classical trajectory is fixed by the Hamilton-Jacobi relation  $p_\mu = \partial_\mu S$.  Solving the on-shell condition $p^2 = 0$ with the ansatz $p^\mu = (\omega, p_r)_{t,r}$, we obtain two roots: 
\begin{equation}
	p^{(1)}_r = -\omega, \qquad p^{(2)}_r = \frac{2 \Gn M\omega + r\omega}{r-2 \Gn M}.
\end{equation}

The first solution corresponds to the classical picture, where no radial radiation can occur.  The second solution hints at vacuum decay, admitting a simple pole at $r=2 \Gn M$. Performing the contour integration around the pole gives rise to the imaginary part of the action, yielding
\begin{equation}
	P^{\text{GR}}_{\text{pair}} \sim \exp \left( - 8 \pi \Gn M \omega \right) \, .
\end{equation}
The horizon pole in gravity thus plays precisely the role of the dual color pole at $r=0$ in the Yang-Mills root: in both cases the imaginary part of the action is fixed by a topological winding around a simple pole of the canonical momentum.

\bibliographystyle{JHEP}
\bibliography{nonperturbDC}

\end{document}